\documentclass[a4paper,12pt]{article}
\usepackage[centertags]{amsmath}
\usepackage{amsfonts}
\usepackage{amssymb}
\usepackage{amsthm}
\usepackage{newlfont}
\usepackage{epsfig}
\usepackage{amscd}

\theoremstyle{plain}
\newtheorem{Th}{Theorem}
\newtheorem{Cor}[Th]{Corollary}
\newtheorem{Lem}[Th]{Lemma}
\newtheorem{Prop}[Th]{Proposition}
\theoremstyle{definition}
\newtheorem{Def}{Definition}

\theoremstyle{remark}
\newtheorem*{Rem}{Remark}
\newcommand{\PP}{{\mathbb P}}
\newcommand{\RR}{{\mathbb R}}
\newcommand{\EE}{{\mathbb E}}
\newcommand{\ZZ}{{\mathbb Z}}

\newcommand{\bx}{{\boldsymbol x}}

\newcommand{\bN}{{\boldsymbol N}}
\newcommand{\by}{{\boldsymbol y}}

\newcommand{\bw}{{\boldsymbol w}}
\newcommand{\bz}{{\boldsymbol z}}

\newcommand{\bX}{{\boldsymbol X}}
\newcommand{\bY}{{\boldsymbol Y}}

\newcommand{\vb}{\vec{\boldsymbol b}}
\newcommand{\vB}{\vec{\boldsymbol B}}
\newcommand{\vbx}{\vec{\bx}}
\newcommand{\vby}{\vec{\by}}

\newcommand{\vbX}{\vec{\bX}}
\newcommand{\vbY}{\vec{\bY}}

\newcommand{\vecbfeta}{\vec{\boldsymbol \eta}}
\newcommand{\vbN}{\vec{\bN}}
\newcommand{\vbn}{\vec{\boldsymbol n}}
\newcommand{\vbomega}{\vec{\boldsymbol \omega}}

\begin{document}
\begin{center}
{\Large Geometric discretization of the Bianchi system} 

\bigskip
{\large A. Doliwa$^*$, M. Nieszporski$^{\dag \ddag}$, P. M. Santini$^\S$} 

\bigskip

{\small
$^*${\it Wydzia{\l} Matematyki i Informatyki, Uniwersytet
Warmi\'nsko--Mazurski

ul. \.Zo{\l}nierska 14A, 10-561 Olsztyn, Poland}

e-mail: {\tt doliwa@matman.uwm.edu.pl} 

\bigskip

$^\dag${\it Katedra Metod Matematycznych Fizyki, Uniwersytet Warszawski

ul. Ho\.za 74, 00-681 Warszawa, Poland}
 
e-mail: {\tt maciejun@fuw.edu.pl} 

\bigskip

$^\ddag${\it Instytut Fizyki Teoretycznej, Uniwersytet w Bia{\l}ymstoku

ul. Lipowa 41, 15-424 Bia{\l}ystok, Poland}

\bigskip 

$\S${\it Dipartimento di Fisica, Universit\`a di Roma ,,La Sapienza''
 
Istituto Nazionale di Fisica Nucleare, Sezione di Roma

P-le Aldo Moro 2, I--00185 Roma, Italy}

e-mail: {\tt paolo.santini@roma1.infn.it} 
} 
\end{center}

\begin{abstract}
We introduce the dual Koenigs
lattices, which are the integrable
discrete analogues of conjugate nets with equal tangential
invariants, and we find the
corresponding reduction of the fundamental transformation. 
We also introduce the notion of discrete normal congruences.
Finally, considering quadrilateral lattices "with equal tangential 
invariants" which allow for harmonic normal congruences we obtain,
in complete analogy with the continuous case, the integrable discrete analogue of
the Bianchi system together with its geometric meaning. To obtain this geometric
meaning we also make use of the novel characterization of the circular lattice
as a quadrilateral lattice whose coordinate lines intersect orthogonally in the
mean.\\

{\it 2003 PACS:} 02.30.Ik, 02.40.Dr, 04.20.Jb, 04.60.Nc 

\end{abstract}

\section{Introduction}

The paper concerns with the integrable discrete analogue of the following
nonlinear partial differential equation 
\begin{equation} \label{eq:Bianchi-diff-1}
\vbN_{,uv} = - \frac{\vbN_{,u}\cdot\vbN_{,v}}{U+V} \vbN, \qquad 
\vbN\cdot \vbN = U+V,
\end{equation} 
where $\vbN$ is a vector valued function of two variables $u$ and $v$,
comma denotes differentiation 
(e.g., $\vbN_{,u}=\frac{\partial \vbN}{\partial u}$), and  $U=U(u)$ and 
$V=V(v)$ are 
given functions of single arguments, $u$ and $v$ respectively.
The system \eqref{eq:Bianchi-diff-1} was introduced long time
ago by L.~Bianchi~\cite{Bianchi} as an equation satisfied by a normal vector
of special hyperbolic surfaces in $\EE^3$; here
$(u,v)$ are the asymptotic coordinates of the surface. The second geometric
interpretation of the Bianchi system is given in terms of 
conjugate nets in $\EE^3$ permanent in deformation~\cite{Eisenhart-TCS}. Such nets can be
equivalently characterized as conjugate nets
with equal tangential invariants allowing 
for harmonic normal 
congruences~\cite{Eisenhart-TS}. 

The third and more recent geometric interpretation of the Bianchi system follows
from its equivalence (with the
signature of the scalar product changed to $+--$) to the Ernst-like
reduction of Einstein's equation describing the interaction of 
gravitational waves \cite{ChandrasekharFerrari,Griffiths-CPWGR}. 
Indeed, define~\cite{NDS}
\begin{equation*}
\xi = \frac{N_1+iN_2}{\sqrt{r}+N_0}, \qquad 
r=\vbN\cdot\vbN = N_0^2 + \epsilon (N_1^2 + N_2^2), \qquad \epsilon = \pm 1;
\end{equation*} 
then the Bianchi system is transformed into the equations
\begin{equation*}
\left( 2\xi_{,uv} + \frac{r_{,v}}{r}\xi_{,u} + \frac{r_{,u}}{r}\xi_{,v}\right)
(\xi\bar{\xi} + \epsilon ) = 4\bar{\xi}\xi_{,u}\xi_{,v}, \qquad r_{,uv}=0,
\end{equation*}
which are the hyperbolic version of the Ernst 
system describing
axisymmetric stationary vacuum solutions of the Einstein equations 
\cite{Ernst,Maison,BelinskiZakharov,Neugebauer}.
 
During the last few years the integrable discrete (difference) 
analogues of 
geometrically significant integrable differential equations have attracted
considerable attention 
\cite{BobenkoPinkall-DSIS,MQL,TQL,KoSchief2}. The integrable discrete
systems appear naturally in statistical physics \cite{JM} and in
quantum field
theory \cite{KBI}. 
It has long been expected that a quantization of gravitational systems will lead
to a discrete structure of space-time and then the differential equations must
be replaced by difference equations at a fundamental level
(see recent reviews \cite{KauffmanSmolin,AJL}).
It would be therefore useful and instructive to study the geometric properties of 
integrable discrete version of the Bianchi system, a distinctive
integrable reduction of Einstein's equations.

In a recent work \cite{NDS,DNS-I}, in which the original paper \cite{Bianchi} of
Bianchi was discretized step by step, the following nonlinear vector equation
\begin{equation} \label{eq:Bianchi-disc}
\vbN_{(12)} + \vbN = \frac{U_1 + U_2}
{(\vbN_{(1)} + \vbN_{(2)})\cdot (\vbN_{(1)} + \vbN_{(2)})}
( \vbN_{(1)} + \vbN_{(2)}),
\end{equation}
was derived, in the context of asymptotic lattices, and identified as the proper
integrable discrete analogue of the Bianchi system \eqref{eq:Bianchi-diff-1}. In
equation \eqref{eq:Bianchi-disc}, $U_1=U_1(m_1)$ and $U_2=U_2(m_2)$ are given
functions, respectively, of the single arguments $m_1$ and $m_2$, and the
subscripts in bracket denote shifts in the indexed variables, i.e., for 
$m_1,m_2\in\ZZ^2$ we write
$\vbN_{(\pm1)}(m_1,m_2)= \vbN(m_1\pm1,m_2)$, 
$\vbN_{(\pm2)}(m_1,m_2)= \vbN(m_1,m_2\pm 1)$,
$\vbN_{(\pm 1 \pm 2)}(m_1,m_2)= \vbN(m_1\pm 1,m_2\pm 1)$.

It is remarkable that the same system \eqref{eq:Bianchi-disc} had been already
introduced, a bit earlier, in~\cite{Schief-C}, in the different geometric context
of discrete isothermic nets, as an integrable discretization, instead, of the
vectorial Calapso equation. Therefore the system 
\eqref{eq:Bianchi-disc} can be viewed as a remarkable example in which
discretization leads to a unification of different geometries and of different
partial differential equations.

Equation \eqref{eq:Bianchi-disc} can be obtained from the discrete Moutard
equation found in \cite{NiSchief}
\begin{equation} \label{eq:Moutard-disc}
\vbN_{(12)} + \vbN = F( \vbN_{(1)} + \vbN_{(2)}),
\end{equation}
imposing the integrable quadratic constraint
\begin{equation} \label{eq:constraint-disc}
(\vbN_{(12)} + \vbN) \cdot( \vbN_{(1)} + \vbN_{(2)}) = U_1 + U_2.
\end{equation}
The integrability of the discrete Bianchi system \eqref{eq:Bianchi-disc} was
proven in~\cite{Schief-C} and \cite{DNS-I} using different approaches, 
by showing
the compatibility of the constraint \eqref{eq:constraint-disc} with a suitable
restriction of the Darboux transformation of the discrete Moutard equation 
\eqref{eq:Moutard-disc}. However the geometric meaning of the constraint 
\eqref{eq:constraint-disc} was still unclear both 
in the context of discrete isothermic nets and in the context of
discrete asymptotic nets. This situation
seemed to be in contrast to the believe that geometry,
and especially the discrete geometry, should
help to understand integrability of the underlying systems. 

In the present paper we obtain the integrable discrete analogue of the Bianchi
system \eqref{eq:Bianchi-diff-1} in a geometric way. In order to keep track of
integrability on each step of the construction, we make use of its geometric
interpretation as the system describing
{\em conjugate nets with equal tangential invariants allowing 
for harmonic normal 
congruences} \cite{Eisenhart-TS}. 

The main results of the paper can be stated as follows: i) we have introduced 
the notion of {\em dual
Koenigs lattice}, the proper integrable
discrete analogue of a conjugate net with equal
tangential invariants; ii) we have introduced the notion of {\em discrete normal
congruence}, the proper integrable
discrete analogue of a normal congruence; iii) we have
introduced the (quadrilateral) Bianchi lattice as a {\em dual Koenigs lattice 
allowing
for a harmonic congruence which is a discrete normal congruence}. 
Because the
simultaneous application of several integrable constraints
preserves integrability, we obtain for free the integrability of the Bianchi
lattice. An additional result of the paper, relevant in deriving the above
results, is the novel geometric characterization of the circular lattice as a
quadrilateral lattice whose coordinate lines intersect orthogonally in the mean.

It turns out that applying the above geometric constraints to the quadrilateral
lattice one obtains 
\begin{equation} \label{eq:Bianchi-Koenigs-disc}
\vbn_{(12)} + \vbn  = F_{(1)} \vbn_{(1)} + 
F_{(2)}  \vbn_{(2)},\qquad
\vbn\cdot\vbn \: F  = U_{1} + U_2,
\end{equation}
which is simply related to equations 
\eqref{eq:Moutard-disc}-\eqref{eq:constraint-disc} by
\[
\vbn = \vbN_{(1)} + \vbN_{(2)}.
\]
In fact the system 
\eqref{eq:Bianchi-Koenigs-disc} appeared 
first
in a hidden form in \cite{NDS} (see also \cite{DNS-I}) as a usefull tool to
prove the integrability of equation \eqref{eq:Bianchi-disc}.   
The program realized in this paper allows to embed yet another integrable
discrete system into the general theory of
lattices with planar quadrilaterals \cite{DCN,MQL,CDS,DMS,KoSchief2,TQL}, which
are the proper discrete analogues of conjugate nets. Moreover, the paper 
confirms
once more that, in the context of integrable geometries, the transition
{\em from differential to discrete}, which is often highly nontrivial on the
a level of the direct calculations, can be simplified and better understood on
geometric level.

The paper is constructed as follows. 
We start from Section \ref{sec:Omega}, in which we present
the integrable discrete analogues of nets with equal tangential invariants,
combining the notion of dual lattice
\cite{DS-sym} with the notion of Koenigs lattice \cite{Dol-Koe}. In Section
\ref{sec:normal-cong} we introduce and study the discrete
normal congruences. Finally, in Section \ref{sec:Bianchi} we put all the above
ingredients together to get the integrable discrete Bianchi system. The
Appendix contains additional results of more algebraic nature stated in the
language of partial difference equations. In its first part
we present the algebraic version of the geometric characterization of
circular lattices and normal congruences obtained in 
Section~\ref{sec:normal-cong}. The second part of the Appendix contains the
Darboux transformation for the discrete Bianchi 
system~\eqref{eq:Bianchi-Koenigs-disc} written as a linear problem.

We remark that, during the conference in which the geometric
discretization described in this paper was presented, also the
discretization of the notion of conjugate net invariant for deformation
was presented \cite{Schief-Arran}. Therefore two of the geometric meanings 
of the Bianchi system \eqref{eq:Bianchi-diff-1} seem to be now successfully 
discretized.

\section{The dual Koenigs lattices}
\label{sec:Omega}
As it was mentioned above, in the geometric derivation of the
Bianchi system~\eqref{eq:Bianchi-diff-1} one combines the notion of the
net with equal tangential invariants with the notion of the normal
congruence. This Section is devoted to the presentation of the integrable
discrete analogue of the first component of the geometric definition of the
Bianchi system. 
We start with collecting some by now classical results from the theory of
two dimensional quadrilateral lattices in three dimensional space 
\cite{Sauer,DCN,MQL}, together with their tangential description
\cite{DS-sym}, and from some recent results on the theory of Koenigs lattices 
(the discrete analogues of conjugate nets
with equal point invariants)~\cite{Dol-Koe}. 

\subsection{Quadrilateral lattices and discrete congruences}
The integrable discrete analogues of conjugate nets are lattices made out of
planar quadrilaterals \cite{Sauer,DCN}, called in \cite{MQL} the
quadrilateral lattices.
\begin{figure}
\begin{center}
\leavevmode\epsfysize=3cm\epsffile{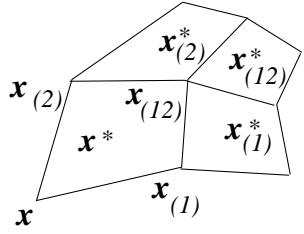}
\end{center}
\caption{The quadrilateral lattice}
\label{fig:sczp-spczp}
\end{figure}
Given such a lattice $x:\ZZ^2\to\PP^3$,
then, in terms of the homogeneous coordinates $\bx:\ZZ^2\to\RR^{4}_*$ of the 
lattice, the planarity of the elementary quadrilaterals can be 
expressed as a linear relation between $\bx$, $\bx_{(1)}$, $\bx_{(2)}$ 
and $\bx_{(12)}$. In the generic situation, assumed in the sequel,
in which three vertices of the elementary quadrilaterals are never collinear, 
the linear relation can be
written down as the discrete Laplace equation~\cite{DCN,MQL} 
\begin{equation} \label{eq:Laplace-dis-gen-2}
 \bx_{(12)} = A_{(1)}\bx_{(1)} + B_{(2)}\bx_{(2)} + C\bx,
\end{equation}
where $A$, $B$ and $C$ are real functions on $\ZZ^2$. 

Any quadrilateral lattice in $\PP^3$ 
can be considered as
envelope of its tangent planes (the planes of elementary quadrilaterals).
Denote by $\bx^*  \in \RR^4_*$ the homogeneous coordinates of the plane
$x^*  \in (\PP^3)^*$ passing through $x$, $x_{(1)}$ and $x_{(2)}$ 
(see Figure \ref{fig:sczp-spczp}), i.e.,
\[  \langle \bx^*, \bx \rangle = \langle \bx^*, \bx_{(1)} \rangle =
\langle \bx^*, \bx_{(2)} \rangle =0. \]
Because the planes $x^*$, $x^*_{(1)}$, $x^*_{(2)}$ and
$x^*_{(12)}$ intersect in the point $x_{(12)}$, then also the homogeneous 
coordinates $\bx^*:\ZZ^2\to\RR^4_*$ satisfy the Laplace equation 
\begin{equation} \label{eq:Laplace-dis-gen-2-dual}
\bx_{(12)}^* = A^*_{(1)} \bx^*_{(1)} + B^*_{(1)} \bx^*_{(2)} + C^* \bx^*.
\end{equation}
The homogeneous coordinates $\bx$ are called the point coordinates of the
lattice $x$ while $\bx^*$ are called the tangential coordinates of the
lattice (see also \cite{DS-sym}).

\begin{figure}
\begin{center}
\leavevmode\epsfysize=4cm\epsffile{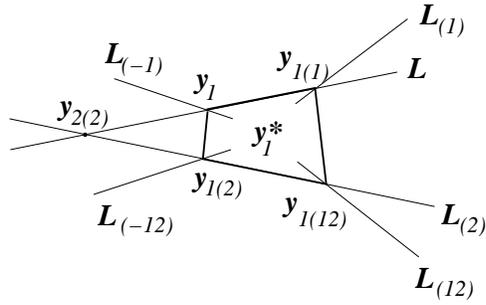}
\end{center}
\caption{The first focal lattice of a congruence}
\label{fig:focal-sczp}
\end{figure}

The theory of transformations of the quadrilateral lattices is based
on the theory of congruences of lines \cite{TQL}.
A $\ZZ^2$ family of straight lines $L$
in $\PP^3$ is called {\em congruence} if any
two neighbouring lines are coplanar (and therefore intersect). 
The intersection points
$y_i=L\cap L_{(-i)}$, $i=1,2$, define focal lattices of the 
congruence. It
turns out that the focal
lattices have planar quadrilaterals as well 
(see Figure~\ref{fig:focal-sczp}). 
\begin{Cor} \label{cor:dual-changes}
Notice, that the plane $y_1^*$ of the first focal lattice 
contains the lines
$L$ and $L_{(2)}$, while the point $y_1$ is the intersection of the
lines $L$ and $L_{(-1)}$. This implies that one cannot just 
"dualize" formulas where focal lattices appear. In general, in place of
$y_1$ it should be written $y^*_{2(-1)}$ while in place of
$y_2$ it should be written $y^*_{1(-2)}$, see Figure~\ref{fig:focal-sczp}. 
\end{Cor}
A quadrilateral lattice and a congruence are said to be {\em conjugate} to
each other if the points of the lattice belong to the corresponding lines 
of the congruence. A quadrilateral lattice and a congruence are said to 
be {\em harmonic} to
each other if the lines of the congruence belong to the corresponding tangent 
planes of the lattice. 
Under the standard dualization in the projective space
$\PP^3$ (see, e.g.,  \cite{Samuel}), the statement "the point belongs to
the line" is replaced by the statement "the plane contains the line". 
This implies that the notions of conjugate and harmonic congruences (to a
quadrilateral lattice) are dual to each other. 

\begin{figure}
\begin{center}
\leavevmode\epsfysize=6cm\epsffile{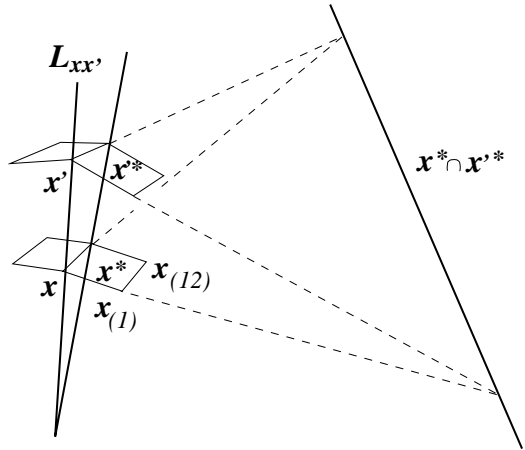}
\end{center}
\caption{The fundamental transformation}
\label{fig:fund-dual}
\end{figure}

The quadrilateral lattice $x^\prime$ is a 
{\em fundamental transform} of $x$ if both lattices are conjugate 
to the same congruence \cite{TQL} (see Figure \ref{fig:fund-dual}), called
the conjugate congruence of the transformation. 
It turns out that both lattices $x$ and $x^\prime$ are also harmonic to 
the congruence given by the intersection of the planes $x^*$ and $x^{\prime*}$,
which is called the harmonic congruence of the transformation. 
This makes the
geometric picture of the fundamental transformation in $\PP^3$ self-dual.

\subsection{The dual Koenigs lattice} 
\begin{figure}
\begin{center}
\leavevmode\epsfysize=7cm\epsffile{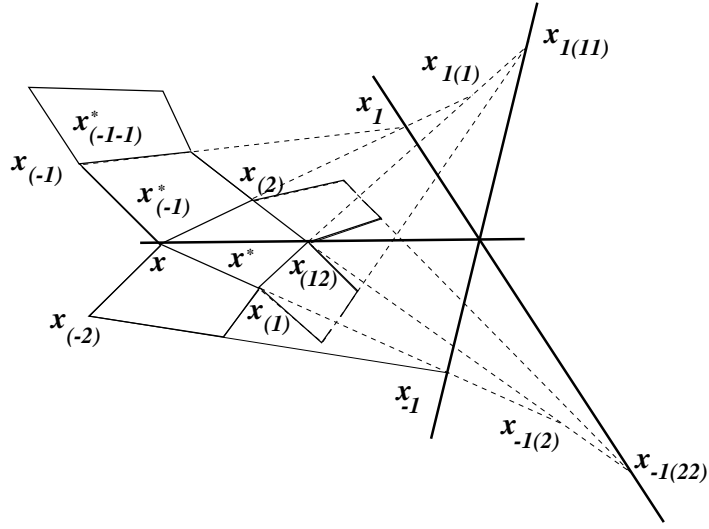}
\end{center}
\caption{The three lines in the definition of the Koenigs lattice}
\label{fig:Lapl-transf-d-K}
\end{figure}

We first recall the geometric definition of a Koenigs lattice and its algebraic
characterization introduced in \cite{Dol-Koe}. 
Given a quadrilateral lattice
$x$ in $\PP^3$, denote by $x_1$ the intersection points of the lines 
$L_{x x_{(2)}}$ with the lines $L_{x_{(-1)} x_{(-12)}}$ and denote by  
$x_{-1}$ the intersection points of the lines 
$L_{x x_{(1)}}$ with the lines $L_{x_{(-2)} x_{(1-2)}}$.
\begin{Def}
{\em The Koenigs lattice} is a two dimensional quadrilateral lattice such that, 
for any point $x$ of the lattice, there exists a conic passing through the six
points $x_{1}$, $x_{1(1)}$,$x_{1(11)}$, 
$x_{-1}$, $x_{-1(2)}$ and $x_{-1(22)}$.
\end{Def}
\begin{Prop} \label{prop:alg-Koe}
The Laplace equation of the 
Koenigs lattice can be gauged into the canonical form 
\begin{equation} \label{eq:Koenigs-d}
\bx_{(12)} + \bx =F_{(1)}\bx_{(1)} + F_{(2)}\bx_{(2)}.
\end{equation}
\end{Prop}
Using Pascal's "mystic hexagon" theorem (see, e.g.,  \cite{Samuel}),
it is possible to obtain the following
alternative geometric characterization of a Koenigs lattice (see Figure
\ref{fig:Lapl-transf-d-K}).
\begin{Prop} \label{prop:dKoenigs-def-predual}
The quadrilateral lattice $x$ in $\PP^3$ is a Koenigs lattice
if and only if, for any point $x$ of the lattice, the three lines\\ 
(i) the line connecting the points 
$x^* \cap x_{(-1)}^* \cap x_{(-1-1)}^*$ and
$x^* \cap x_{(2)}^* \cap x_{(22)}^*$,\\ 
(ii) the line connecting the points 
$x^* \cap x_{(-2)}^* \cap x_{(-2-2)}^*$ and
$x^* \cap x_{(1)}^* \cap x_{(11)}^*$, \\
(iii) the line connecting the points 
$x^* \cap x_{(-1)}^* \cap x_{(-2)}^*$ and
$x^* \cap x_{(1)}^* \cap x_{(2)}^*$, \\
intersect in a single point.
\end{Prop} 
\begin{proof}
By the Pascal theorem, the quadrilateral lattice $x$ is a Koenigs lattice 
if and only if the
lines $L_{x_{-1}x_{1(11)}}$ and $L_{x_{1}x_{-1(22)}}$ intersect in a point of 
$L_{x x_{(12)}}$. Notice that in $\PP^3$
the point $x_1$
is the intersection point of the three planes $x^*$, $x_{(-1)}^*$ and
$x_{(-1-1)}^*$. Similarly, the point $x_{-1}$ is the intersection 
point of three planes $x^*$, $x_{(-2)}^*$ and $x_{(-2-2)}^*$. By definition, 
in $\PP^3$ we have $x=x^* \cap x_{(-1)}^* \cap x_{(-2)}^*$ and
$x_{(12)}=x^* \cap x_{(1)}^* \cap x_{(2)}^*$.
\end{proof}
\begin{Rem}
For a generic quadrilateral lattice $x$ in $\PP^3$, the three lines of the
above proposition are contained in the plane $x^*$.
\end{Rem}

The notion of the dual Koenigs lattice can be conveniently obtained by dualizing
the geometric definition of the Koenigs lattice given in the above proposition.
\begin{Def} \label{def:dual-dKoenigs}
A quadrilateral lattice in $\PP^3$ is called the 
{\em dual Koenigs lattice}
if, for any point $x$ of the lattice, the three lines\\  
(i) the intersection line of the plane
$\pi_{x x_{(-1)} x_{(-1-1)}}$ with the plane
$\pi_{x x_{(2)} x_{(22)}}$,\\ 
(ii) the intersection line of the plane
$\pi_{x x_{(-2)} x_{(-2-2)}}$ with the plane
$\pi_{x x_{(1)} x_{(11)}}$,\\ 
(iii) the intersection line of the plane
$\pi_{x x_{(-1)} x_{(-2)}}$ with the plane
$\pi_{x x_{(1)} x_{(2)}}$,\\
are coplanar.
\end{Def} 
\begin{Rem}
For a generic quadrilateral lattice in $\PP^3$ the three lines of the
above definition intersect in the point $x$.
\end{Rem}

Also the basic algebraic characterization of the
dual Koenigs lattice can be obtained  dualizing the corresponding algebraic
characterization of the Koenigs lattice (just replacing 
the homogeneous point
coordinates by the tangential ones). Because of the importance of this
characterization in the paper, we will prove it without referring 
to the duality principle.   
\begin{Prop} \label{prop:alg-Koe-d}
The quadrilateral lattice $x$ in $\PP^3$ is a dual Koenigs
lattice if and only if its tangential coordinates can be gauged in such 
a way that the Laplace equation~\eqref{eq:Laplace-dis-gen-2-dual} takes the form
\begin{equation} \label{eq:Koenigs-dis-dual}
\bx_{(12)}^* + \bx^* = F_{(1)}\bx_{(1)}^* + F_{(2)}\bx_{(2)}^*.
\end{equation}
\end{Prop}
\begin{proof}
The plane $\pi_{x x_{(1)} x_{(11)}}$ contains the lines
$L_{x\,x_{(1)}}=x^*\cap x^*_{(-2)}$ and 
$L_{x_{(1)}\,x_{(11)}}=x_{(1)}^*\cap x^*_{(1-2)}$. Therefore the homogeneous
coordinates of the plane can be found by solving the linear system
\begin{equation*}
\lambda \bx^* + \mu \bx^*_{(-2)} = 
\sigma \bx^*_{(1)} + \delta \bx^*_{(1-2)}. 
\end{equation*}
Using the Laplace
equation~\eqref{eq:Laplace-dis-gen-2-dual} it can be shown that the 
coordinates read
\begin{equation*}
B^* \bx^* + C^*_{(-2)} \bx^*_{(-2)} = 
\bx^*_{(1)} -A^*_{(1-2)} \bx^*_{(1-2)}. 
\end{equation*}
Similarly, the homogeneous coordinates of the planes 
$\pi_{x x_{(2)} x_{(22)}}$, $\pi_{x x_{(-1)} x_{(-1-1)}}$ and
$\pi_{x x_{(-2)} x_{(-2-2)}}$ are given, correspondingly, by
\begin{eqnarray}
A^* \bx^* + C^*_{(-1)} \bx^*_{(-1)} & = &
\bx^*_{(2)} -B^*_{(-12)} \bx^*_{(-12)}, \nonumber \\
B^*_{(-1-1)} \bx^*_{(-1-1)} + C^*_{(-1-1-2)} \bx^*_{(-1-1-2)} & = &
\bx^*_{(-1)} -A^*_{(-1-2)} \bx^*_{(-1-2)},  \nonumber \\
A^*_{(-2-2)} \bx^*_{(-2-2)} + C^*_{(-1-2-2)} \bx^*_{(-1-2-2)} & = &
\bx^*_{(-2)} -B^*_{(-1-2)} \bx^*_{(-1-2)}  \nonumber
\end{eqnarray}
The coordinates of the planes $\pi_{x x_{(1)} x_{(2)}}$ and
$\pi_{x x_{(-1)} x_{(-2)}}$ are, by definition, $\bx^*$ and 
$\bx^*_{(-1-2)}$. 
There exists a plane containing the intersection lines 
$\pi_{x x_{(1)} x_{(11)}}\cap \pi_{x x_{(-2)} x_{(-2-2)}}$, 
$\pi_{x x_{(2)} x_{(22)}}\cap \pi_{x x_{(-1)} x_{(-1-1)}}$ and
$\pi_{x x_{(1)} x_{(2)}}\cap \pi_{x x_{(-1)} x_{(-2)}}$ if and only if
the linear system for the unknowns $\lambda,\mu,\sigma,\delta,\chi,\nu$
\begin{gather*}
\lambda (B^* \bx^* + C^*_{(-2)} \bx^*_{(-2)}) + 
\mu(\bx^*_{(-2)} -B^*_{(-1-2)} \bx^*_{(-1-2)}) = \\
\sigma(A^* \bx^* + C^*_{(-1)} \bx^*_{(-1)}) +
\delta(\bx^*_{(-1)} -A^*_{(-1-2)} \bx^*_{(-1-2)}) =
\chi \bx^* + \nu \bx^*_{(-1-2)},
\end{gather*}
has a non-trivial solution. By linear algebra, and assuming that no three of
the four points $x$, $x_{(-1)}$, $x_{-2)}$ and $x_{(-1-2)}$ are collinear,
such a solution exists when 
\begin{equation*}
A^* C^*_{(-2)}B^*_{(-1-2)} = B^* C^*_{(-1)}A^*_{(-1-2)}.
\end{equation*}
This restriction on the coefficients of the Laplace
equation~\eqref{eq:Laplace-dis-gen-2-dual} implies, exactly like in the
corresponding proposition of \cite{Dol-Koe}, existence of the gauge function
$\rho$ defined by
\begin{equation*}
\rho_{(12)} = -C^*\rho, \qquad \rho_{(1)} A^* = \rho_{(2)}B^*.
\end{equation*}
After the gauge transformation $\bx^*\mapsto\bx^* /\rho$, the new homogeneous
tangential coordinates of the lattice satisfy the Laplace equation of the
form~\eqref{eq:Koenigs-dis-dual} with the potential
\begin{equation*}
F=\frac{A^*\rho}{\rho_{(2)}} = \frac{B^*\rho}{\rho_{(1)}}.
\end{equation*}
\end{proof}
After having established the relation between Koenigs lattices and dual
Koenigs lattices on both geometric and algebraic levels, we will make use of
the results of \cite{Dol-Koe} to present the 
dual version of other basic properties of the Koenigs lattices.
Taking into account the exchange of 
indexes and shifts in the notation for the tangential coordinates of the focal
lattices of the harmonic congruence (see Corollary~\ref{cor:dual-changes}), 
we will obtain the geometric characterization of the dual Koenigs lattices
from the point of view of their transformations.

Consider a quadrilateral lattice $x$ in $\PP^3$ and a 
congruence $L$ harmonic to the lattice. According to the dual description of
lines, any line of the congruence is identified with
the pencil of planes containing the line. Notice that the planes 
$z_1^*$, $z_2^*$, $z_{1(-2)}^*$ and $z_{2(-1)}^*$ of the focal lattices of the
congruence, as well as the plane $x^*$, are elements of the same pencil. 
On each line there exists a unique involution, denoted by ${\mathfrak i}$,
which maps the planes $z_1^*$ and $z_2^*$ to the planes
$z_{1(-2)}^*$ and $z_{2(-1)}^*$, 
correspondingly. The following proposition is the dual version of the
Corollary~15 of \cite{Dol-Koe}.
\begin{Prop} \label{prop:invol-harm-ddK}
The quadrilateral lattice $x$ in $\PP^3$ is a dual Koenigs lattice
if and only if, for any congruence harmonic to the lattice, the planes
$x^{*\prime}={\mathfrak i}(x^*)$, where 
${\mathfrak i}$ are the involutions described above, 
are tangent planes of a quadrilateral lattice.
\end{Prop} 
\begin{Cor}
If $x$ is a dual Koenigs lattice, then the quadrilateral
lattice $x^\prime$ with tangent planes $x^{\prime*}={\mathfrak i}(x^*)$ 
must be, by the symmetry of the construction, a
dual Koenigs lattice as well. 
\end{Cor}
The above Proposition provides the basic geometric characterization of the
reduction of the fundamental transformation corresponding to the dual Koenigs
lattices. 
Its algebraic formulation, given by the dual version of Proposition~8 of
\cite{Dol-Koe}, reads as follows; here $\Delta_i$, $i=1,2$, denote the 
standard partial difference operators acting as
$\Delta_i f= f_{(i)}-f$. 
\begin{Th} \label{th:tr-dKoe-du}
Given a dual Koenigs lattice with tangential coordinates $\bx^*$
in the ca\-no\-ni\-cal gauge of equation \eqref{eq:Koenigs-dis-dual}.
If $\theta$ is a scalar solution 
of the discrete Moutard equation 
\begin{equation} \label{eq:Moutard-dis}
\theta_{(12)}+\theta = F(\theta_{(1)}+\theta_{(2)}),
\end{equation}
then the solution $\bx^{\prime *}$ of the linear system
\begin{align} 
\Delta_1\left( \frac{\bx^{\prime *}}{\phi^\prime} \right) & = \; \; \; 
 \theta_{(1)} \theta_{(12)} \Delta_1\left( \frac{\bx^*}{\phi} \right), \\
\Delta_2\left( \frac{\bx^{\prime *}}{\phi^\prime} \right) & = 
- \theta_{(2)} \theta_{(12)}\Delta_2\left( \frac{\bx^*}{\phi} \right),
\end{align}
with $\phi$ and $\phi^\prime$ given by 
\begin{equation} \label{eq:phi-phi'}
\phi = \theta_{(1)} + \theta_{(2)}, \quad \phi^\prime = 
 \frac{1}{\theta_{(1)}} + \frac{1}{\theta_{(2)}},
\end{equation}
gives the tangential coordinates of a new dual Koenigs lattice satisfying 
equation 
\eqref{eq:Koenigs-dis-dual}, with the potential 
\begin{equation} \label{eq:F'}
F^\prime=F\frac{\theta_{(1)}\theta_{(2)}}{\theta \:\theta_{(12)}}.
\end{equation}
\end{Th}
Finally, we collect the dual versions of some formulas scattered in \cite{Dol-Koe}
which turn out to be useful in the sequel.
\begin{Cor} \label{cor:dKoe-du-inv}
The homogeneous coordinates of the
tangent planes of the focal lattices $z_{i}$, $i=1,2$
of the harmonic congruence of the dual
discrete Koenigs transformation are given by
\begin{align} \label{eq:focal-lat-Koe-dual-1}
\bz_1^*  & = - \left( \theta\theta_{(1)} \frac{\bx^*}{\phi} + 
\frac{\bx^{\prime*}}{\phi^{\prime}} \right)_{(2)}  =  
-\theta_{(2)}\theta_{(12)} \frac{\bx^*}{\phi} - 
\frac{\bx^{\prime*}}{\phi^\prime} , \\
\label{eq:focal-lat-Koe-dual-2}
\bz_2^* & = \; \; \; \left( \theta\theta_{(2)} \frac{\bx^*}{\phi} - 
\frac{\bx^{\prime*}}{\phi^\prime} \right)_{(1)}  = 
\; \; \;\theta_{(1)}\theta_{(12)} \frac{\bx^*}{\phi} - 
\frac{\bx^{\prime*}}{\phi^\prime} .
\end{align} 
They can be found from equations
\begin{align} \label{eq:z12-x}
\bz_{2(-1)}^* - \bz_{1(-2)}^*  &= \; \; \; \theta \bx^* ,\\
\label{eq:D1z1-2}
\Delta_1\bz_{1(-2)}^*  &=  -\theta_{(1)}(\bx^*_{(1)} - F\bx^*), \\
\label{eq:D2z2-1}
\Delta_2\bz_{2(-1)}^*  &=  \; \; \; \theta_{(2)}(\bx^*_{(2)} - F\bx^*), 
\end{align}
once the solution $\theta$ of equation \eqref{eq:Moutard-dis} is given. 
Moreover, the homogeneous coordinates of the
fixed planes of the involutions ${\mathfrak i}$,  are given by
\begin{equation} \label{eq:fixed-x-xK-dual}
\bz_\pm^*=\pm\sqrt{\theta\theta_{(12)}}\bx^* + 
\sqrt{\theta_{(1)}\theta_{(2)}}\bx^{\prime *}. 
\end{equation}
\end{Cor}

\section{The discrete normal congruences}
\label{sec:normal-cong}
The classical examples \cite{Eisenhart-TCS}
of congruences are given by normals to a surface in the
three dimensional Euclidean space $\EE^3$. The developables of such 
congruences cut the surface along the curvature lines and the corresponding
tangent planes of its focal nets intersect orthogonally. Keeping the last
property one obtains the notion of the normal congruence. 
In this section we will
define the discrete analogue of such congruences.

The discrete analogues of surfaces parametrized by curvature lines were
introduced in \cite{2dcl1,2dcl2} and are two dimensional lattices in
$\EE^3$ made of circular quadrilaterals. 
Integrability of multidimensional circular lattices was proven first
geometrically in \cite{CDS} (see also \cite{Bobenko-O}), and then by 
using analytic means in \cite{DMS}.

\begin{figure}
\begin{center}
\leavevmode\epsfysize=5cm\epsffile{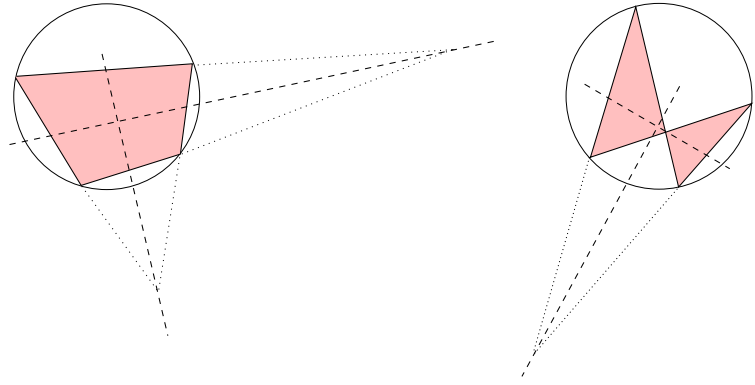}
\end{center}
\caption{Circular quadrilaterals}
\label{fig:dis-curv}
\end{figure}

Let us present a new elementary geometric characterization of circular lattices
which will be important in the characterization of the Bianchi lattices
(for other algebraic characterizations of circular lattices~\cite{CDS,KoSchief2,q-red}
see the Appendix).
\begin{Prop} \label{prop:circ-bis}
A planar quadrilateral is circular 
if and only if the lines bisecting the angles between
its opposite sides and intersecting the quadrilateral 
are orthogonal (see Figure~\ref{fig:dis-curv}).
\end{Prop}
\begin{proof}
By elementary geometry, using the following facts: 
i) a convex planar quadrilateral is
circular if and only if its opposite angles add to $\pi$; 
ii) a non-convex planar quadrilateral is
circular if and only if its opposite angles are equal.
\end{proof}
\begin{Rem}
The above result means that the coordinate lines of the circular lattice 
intersect orthogonally "in the mean".
\end{Rem}

\begin{figure}
\begin{center}
\leavevmode\epsfysize=7cm\epsffile{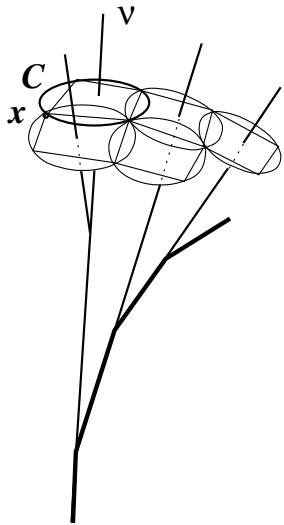}
\end{center}
\caption{Normal congruence of a circular lattice}
\label{fig:dis-norm-cong}
\end{figure}

In \cite{Dol-Rib} it was shown geometrically that the normals to
the circles of a two dimensional circular lattice in $\EE^3$ 
passing through their centers form a congruence. Let us recall that 
construction. Denote by $C$ the circle passing through $x$, $x_{(1)}$ 
and  $x_{(2)}$,
and by $\nu$ denote the line normal to the plane of the
circle and passing through its center (see Figure~\ref{fig:dis-norm-cong}).
The plane bisecting orthogonally the segment $xx_{(2)}$ is the common plane of 
$\nu$ and $\nu_{(-1)}$, and the plane bisecting orthogonally the segment 
$xx_{(1)}$ is the common plane of $\nu$ and $\nu_{(-2)}$, which shows that 
the family of normals $\nu$ is a congruence, called the normal congruence of the
circular lattice. This property
provides the discrete analogue of the
basic relation between normals of a surface in $\EE^3$ and
the curvature lines.
Due to 
Proposition~\ref{prop:circ-bis} we have the discrete counterpart of this
characterization of normal congruences as orthogonality of the focal
planes "in the mean".
\begin{Prop} \label{prop:norm-cong-circ}
If the congruence $\nu$ with focal lattices $y_1$ and $y_2$
is a normal congruence of the circular lattice $x$,
then the pair of orthogonal planes bisecting the angles between
the planes $y_{1(-2)}^*$ and $y_1^*$ coincides with those bisecting  
$y_{2(-1)}^*$ and $y_2^*$ .
\end{Prop}
\begin{figure}
\begin{center}
\leavevmode\epsfysize=6cm\epsffile{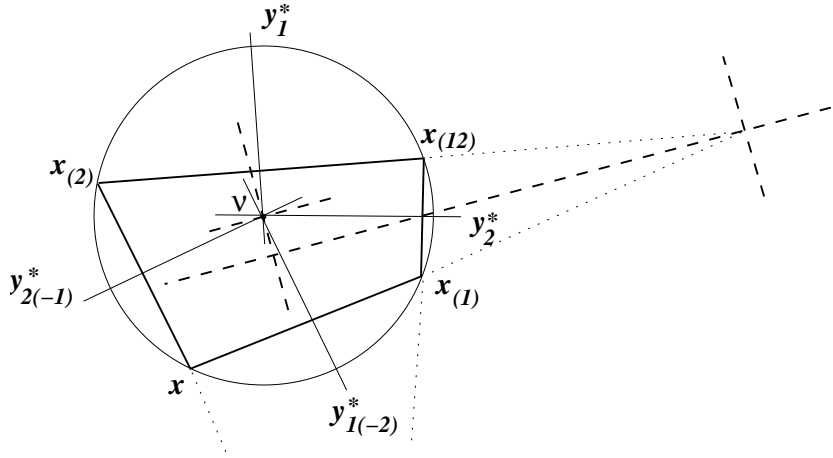}
\end{center}
\caption{Discrete normal congruence}
\label{fig:norm-cong-circ}
\end{figure}
\begin{proof}
Notice that
the plane $y_{1(-2)}^*$ containing the lines $\nu$ and $\nu_{-2}$ is orthogonal
to the segment $xx_{(1)}$. The rest of proof follows from similar 
facts for other planes intersecting at
$\nu$ and other sides of the quadrilateral and from
Proposition~\ref{prop:circ-bis} (see Figure~\ref{fig:norm-cong-circ}).
\end{proof}
The property of congruences normal to circular lattices described above suggests
the following definition of discrete normal congruences.
\begin{Def} \label{def:norn-cong}
A two dimensional congruence in $\EE^3$ is called {\em normal congruence}  
if the pairs of planes bisecting the angles between
its two corresponding pairs of the consecutive focal planes coincide.
\end{Def}

Recall that, within all projective involutions in a pencil of planes in
$\EE^3$,
the reflections are characterized by the property that the fixed planes of the
involution are orthogonal. This gives the 
following, important for further purposes, 
geometric characterization of the normal congruences
\begin{Prop} \label{prop:circ-cong-ref}
Consider a two dimensional congruence in the Euclidean space $\EE^3$ whose
focal lattices have the tangent planes $y^*_i$, $i=1,2$. On each line of the
congruence, consider the unique involution in the corresponding pencil of
planes mapping $y^*_{1(-2)}$ to $y^*_1$ and $y^*_{2(-1)}$ to $y^*_2$. The
congruence is normal if and only if the fixed planes of the involution are
orthogonal. 
\end{Prop}

There is a natural question if all discrete normal congruences can 
be constructed from  
circular lattices. The answer is affirmative but, because we will not use that
fact in the sequel, we present here only a geometric sketch of its proof
(the interested Reader can find in the Appendix
the algebraic description of normal congruences and the algebraic proof
of the above result). 
Start with a point $x\in\EE^3$ and define
its image $x_{(1)}$ in reflection with
respect to the plane $y_{1(-2)}^*$ (see Figure~\ref{fig:norm-cong-circ}).
Similarly, define the point $x_{(2)}$ as 
the image of $x$ in reflection with respect to the plane
$y_{2(-1)}^*$. The construction turns out to be compatible, i.e., 
the image of $x_{(1)}$ in reflection of the plane $y_{2}^*$ is the same as 
the image of $x_{(2)}$ in reflection of the plane $y_{1}^*$, due to 
the normality of
the congruence. Moreover, the point $x_{(12)}$ defined in this way is
concircular with $x$, $x_{(1)}$ and $x_{(2)}$.
Therfore, given the starting point $x_0$, one can construct the
circular lattice with the congruence being its normal congruence.

\section{The discrete Bianchi system}
\label{sec:Bianchi}

The geometric interpretation of the
Bianchi system is given by conjugate nets with equal
tangential invariants allowing for a harmonic congruence which is a normal
congruence \cite{Eisenhart-TS}. In such a situation, the involution in 
the pencil of planes with the base being the harmonic congruence
is the orthogonal reflection with respect to the planes of the focal nets.
In the previous sections we have defined the integrable discrete analogues of the
dual Koenigs net and of the normal congruence. We may therefore expect
that, composing both notions in exactly the same way like in the
continuous case, we should obtain the
discrete Bianchi system together with its geometric interpretation. 
\begin{Def}
The {\em quadrilateral Bianchi lattice} is a dual Koenigs lattice in 
$\EE^3$ allowing 
for a harmonic congruence which is a normal congruence.
\end{Def}
\begin{Rem}
We use the name {\em quadrilateral} Bianchi lattice to distinguish it from the
asymptotic Bianchi lattice described in \cite{DNS-I}. From now on we study only
quadrilateral Bianchi lattices skipping the adjective "quadrilateral".
\end{Rem}
In studying Bianchi lattices we will often restrict the homogeneous
representation of planes to its affine part, i.e., if $\bw^*=(\vbomega,w^{*4})$ is
the tangential homogeneous representation of the plane $w^*$, its 
affine part is $\vbomega$. Notice that in this description all parallel 
planes are indistinguishable.
\begin{Prop} \label{prop:dBianchi-alg}
Let $x$ be a dual Koenigs lattice in $\EE^3$ with tangential coordinates 
$\bx^*$ satisfying equation \eqref{eq:Koenigs-dis-dual}, denote by $\vbn$ its
affine part. 
If the lattice $x$ is a Bianchi lattice, then
\begin{equation} \label{eq:dBianchi-ddKoenigs}
\Delta_1\Delta_2 (\vbn\cdot\vbn \: F) = 0. 
\end{equation}
\end{Prop}
\begin{proof}
Let $L$ be a normal congruence harmonic to $x$. By 
Proposition~\ref{prop:invol-harm-ddK}, we have a unique new dual Koenigs lattice
$x^\prime$, harmonic to the congruence and related with $x$ by the 
corresponding
restriction of the fundamental transformation. Denote by $\vbn^\prime$ the
affine part of the tangential coordinates of $x^\prime$ in the canonical gauge
of Theorem~\ref{th:tr-dKoe-du}. Denote by $\vbn_\pm$ the
affine parts of the tangential coordinates of the fixed planes of the 
involution on $L$
in the gauge of Corollary~\ref{cor:dKoe-du-inv}, i.e.,
\begin{equation} \label{eq:fixed-x-xK-dual-af}
\vbn_\pm =\pm\sqrt{\theta\theta_{(12)}}\vbn + 
\sqrt{\theta_{(1)}\theta_{(2)}}\vbn^\prime . 
\end{equation}
By Proposition~\ref{prop:circ-cong-ref}, the congruence $L$ is normal if and only
if the vectors $\vbn_\pm$ are orthogonal, which gives
the following constraint between the affine parts of the tangential
coordinates of the Bianchi lattice and of its transform
\begin{equation}\label{eq:constraint-dis-1}
\vbn\cdot \vbn \: F = \vbn^{\prime} \cdot \vbn^{\prime} \: F^\prime .
\end{equation}
Denote by $\vbn_i$, $i=1,2$, the
affine parts of the tangential coordinates of the focal lattices  
$z_i$, $i=1,2$, of the
congruence $L$ in the gauge of Corollary~\ref{cor:dKoe-du-inv}, i.e.,
\begin{align} \label{eq:focal-lat-Koe-dual-1-af}
\vbn_1  &=  
-\theta_{(2)}\theta_{(12)} \frac{\vbn}{\phi} - 
\frac{\vbn^{\prime}}{\phi^\prime} = - 
\left( \theta\theta_{(1)} \frac{\vbn}{\phi} + 
\frac{\vbn^{\prime}}{\phi^\prime}\right)_{(2)}, \\
\label{eq:focal-lat-Koe-dual-2-af}
\vbn_2  & = 
\; \; \theta_{(1)}\theta_{(12)} \frac{\vbn}{\phi} - 
\frac{\vbn^{\prime}}{\phi^\prime} = \; \; 
\left( \theta\theta_{(2)} \frac{\vbn}{\phi} - 
\frac{\vbn^{\prime}}{\phi^\prime}\right)_{(1)}.
\end{align} 
Then equation \eqref{eq:constraint-dis-1} implies that 
\begin{equation} \label{eq:squares-n_1n_2n}
\frac{\vbn_{2}\cdot \vbn_{2}}{\theta_{(1)}\theta_{(12)}} +
\frac{\vbn_{1}\cdot \vbn_{1}}{\theta_{(2)}\theta_{(12)}} = \vbn\cdot\vbn \: F,
\end{equation}
and
\begin{equation}
\Delta_1 \left( \frac{\vbn_{2(-1)}\cdot 
\vbn_{2(-1)} }{\theta\theta_{(2)}} \right) =0, \qquad 
\Delta_2 \left( \frac{\vbn_{1(-2)}\cdot 
\vbn_{1} }{\theta\theta_{(1(-2))}} \right) =0, 
\end{equation}
which give the constraint \eqref{eq:dBianchi-ddKoenigs}.
\end{proof}

\begin{Cor}
The condition \eqref{eq:dBianchi-ddKoenigs} implies that
\begin{equation} \label{eq:dBianchi-ddKoenigs-U}
\vbn\cdot\vbn \: F = U_1(m_1) + U_2(m_2),
\end{equation}
where $U_1$ and $U_2$ are functions of the single variables $m_1$ and $m_2$,
respectively. In the above notation
\begin{equation} \label{eq:ni*ni}
\vbn_{1(-2)}\cdot \vbn_{1(-2)}  = \theta\theta_{(1)}
\left(U_1(m_1) + \lambda\right), \quad
\vbn_{2(-1)}\cdot \vbn_{2(-1)}  = \theta\theta_{(2)} 
\left(U_2(m_2) - \lambda\right), 
\end{equation}
where $\lambda$ is a constant.
\end{Cor}
\begin{Cor}
Because the congruence $L$ is also harmonic to the new dual Koenigs lattice
$x^\prime$, then the lattice is also a Bianchi lattice.
\end{Cor}

Our last step is to show that the condition described in 
Proposition~\ref{prop:dBianchi-alg} is also sufficient to characterize the
Bianchi
lattices among the dual Koenigs lattices. In
order to make clear the geometric content of the forthcoming calculations, 
let us
first draw some consequences of the previous considerations.

Let $x$ be a dual Koenigs lattice and let $L$ be a harmonic congruence
conjugated to the lattice. Denote by $\bz^*_1$ and $\bz^*_2$ the tangential
coordinates of the focal lattices of $L$ in the gauge of
Corollary~\ref{cor:dKoe-du-inv}. 
The plane $w^*$ with homogeneous coordinates
\begin{equation} \label{eq:z12-w} \bw^* = -\bz_{1(-2)}^* - \bz_{2(-1)}^*
\end{equation}
contains the line $L$,  while 
equations \eqref{eq:D1z1-2}-\eqref{eq:D2z2-1} imply that
$\bw^*$ satisfies the linear system
\begin{align} \label{eq:lin-w-1}
\Delta_1\bw^* & =  \; \; \theta_{(1)}\bx^*_{(1)} -(2 F\theta_{(1)}-\theta)\bx^*, \\
\label{eq:lin-w-2}
\Delta_2\bw^* & =   -\theta_{(2)}\bx^*_{(2)} +(2 F\theta_{(2)}-\theta)\bx^*,
\end{align}
and, by equations
\eqref{eq:focal-lat-Koe-dual-1}-\eqref{eq:focal-lat-Koe-dual-2}, 
the tangential coordinates of the corresponding Koenigs transform $x^\prime$
read
\begin{equation} \label{eq:tang-x'}
\bx^{\prime *} = \frac{1}{2} \left[ 
\left( \frac{1}{\theta_{(1)}} - \frac{1}{\theta_{(2)}} \right) \theta\bx^* +
\left( \frac{1}{\theta_{(1)}} + \frac{1}{\theta_{(2)}} \right) \bw^* \right].
\end{equation}
If, moreover, $x$ is a Bianchi lattice and the congruence $L$ is normal, 
then the affine
part $\vbomega=-\vbn_{1(-2)}-\vbn_{2(-1)}$ of $\bw^*$ 
is subjected also to the following constraints
\begin{eqnarray}\label{eq:w*n}
\vbomega\cdot\vbn & = & 
\theta_{(1)}(U_1 + \lambda) - \theta_{(2)}(U_2-\lambda) ,\\
\label{eq:w*w}
\vbomega\cdot\vbomega & = & 2\left( (U_1+\lambda)\theta\theta_{(1)} +
(U_2-\lambda)\theta\theta_{(2)}\right) - \theta^2 \frac{U_1+U_2}{F}.
\end{eqnarray}
\begin{Prop}
Let $x$ be a dual Koenigs lattice in $\EE^3$ with tangential coordinates 
$\bx^*$ satisfying equation \eqref{eq:Koenigs-dis-dual}. If its
affine part $\vbn$ satisfies equation~\eqref{eq:dBianchi-ddKoenigs}
then the lattice $x$ is a Bianchi lattice.
\end{Prop}
\begin{proof}
The idea is to construct a normal congruence harmonic to the given dual 
Koenigs lattice subjected to condition~\eqref{eq:dBianchi-ddKoenigs}. 
Let $\bx^*=(\vbn,x^{*4})$ be the homogeneous tangential
coordinates of such a lattice in the canonical gauge. Fix the functions 
$U_1$, $U_2$ in equation~\eqref{eq:dBianchi-ddKoenigs-U} and 
fix a parameter $\lambda$ (it plays the role of spectral parametr in soliton
theory).

Given (from previous steps of the construction or as initial data)
the plane $w^*$ with homogeneous coordinates $\bw^*=(\vbomega,w^{*4})$ and the
scalar $\theta$. Define the planes $z^*_{2(-1)}$ and $z^*_{1(-2)}$ with homogeneous
coordinates $\bz^*_{2(-1)}=(\vbn_{2(-1)},z^{*4}_{2(-1)})$ and
$\bz^*_{1(-2)}=(\vbn_{1(-2)},z^{*4}_{1(-2)})$
by equations
\begin{align} \label{eq:con-z1}
\bz^*_{1(-2)} &= - \frac{1}{2}(\theta \bx^* + \bw^*),\\
\label{eq:con-z2}
\bz^*_{2(-1)} &= \; \; \frac{1}{2}(\theta \bx^* - \bw^*).
\end{align}
Define $\theta_{(1)}$ and $\theta_{(2)}$ by equations \eqref{eq:ni*ni},
then the affine part $\vbomega$ of $\bw^*$ satisfies equations \eqref{eq:w*n}
and \eqref{eq:w*w}. In the last step of the construction we find $\bw^*_{(1)}$
and $\bw^*_{(2)}$ from equations~\eqref{eq:lin-w-1}-\eqref{eq:lin-w-2}.

By equations \eqref{eq:w*n} and \eqref{eq:w*w} the construction is compatible
and, moreover, the function $\theta$ satisfies the discrete Moutard equation 
\eqref{eq:Moutard-dis}. Then equations \eqref{eq:con-z1}-\eqref{eq:con-z2}
and \eqref{eq:lin-w-1}-\eqref{eq:lin-w-2} imply that $\bz^*_{1}$ and $\bz^*_{2}$
satisfy equations \eqref{eq:z12-x}-\eqref{eq:D2z2-1}, i.e., the lattices $z_1$
and $z_2$ are focal lattices of a congruence $L$ harmonic to $x$. 

Because the vectors $\vbn_{1(-2)}/\sqrt{\theta\theta_{(1)}}$ and 
$\vbn_{1}/\sqrt{\theta_{(2)}\theta_{(12)}}$ are of equal length, then the planes
bisecting $z^*_{1(-2)}$ and $z^*_{1}$ have the normal vectors
\begin{equation*}
\vb_{1\mp}=\frac{\vbn_{1(-2)}}{\sqrt{\theta\theta_{(1)}}}\pm
\frac{\vbn_{1}}{\sqrt{\theta_{(2)}\theta_{(12)}}};
\end{equation*}
similarly, the planes
bisecting $z^*_{2(-1)}$ and $z^*_{2}$ have the normal vectors
\begin{equation*}
\vb_{2\pm}=\frac{\vbn_{2(-1)}}{\sqrt{\theta\theta_{(2)}}}\pm
\frac{\vbn_{2}}{\sqrt{\theta_{(1)}\theta_{(12)}}}.
\end{equation*}
Equations~\eqref{eq:dBianchi-ddKoenigs-U}, \eqref{eq:w*n}
and \eqref{eq:w*w} imply 
\begin{equation*}
\vb_{1-} \cdot \vb_{2+} = \vb_{1+} \cdot \vb_{2-} =0,
\end{equation*}
which shows that the congruence $L$ is normal.
\end{proof}
\begin{Cor}
Once the normal congruence $L$ is found then the tangential coordinates of the
corresponding new Bianchi lattice $x^\prime$ are given by \eqref{eq:tang-x'}. 
\end{Cor}
\begin{Rem}
The vectors $\vb_{i\pm}$ are proportional to the vectors
$\vbn_\pm$ considered in the proof
of Proposition~\ref{prop:dBianchi-alg}.
\end{Rem}
\section*{Acknowledgments}
One of the authors (A. D.) would like to thank
dr~J.~Kosiorek for discussion on Proposition~\ref{prop:circ-bis}. A substantial
part of the paper was writen when A. D. was assistant professor at 
the Institute of Theoretical
Physics of Warsaw University.
A.~D. and M.~N. acknowledge partial support from KBN grant no. 2P03B12622. 
We also acknowledge partial support by the cultural
agreement between the University
of Rome "La Sapienza" and the Warsaw University. 

\appendix

\section{Algebraic description of normal congruences}
The goal of the Appendix is to present the theory of discrete 
normal congruences using the more algebraic language of difference equations. 

\subsection{Quadrilateral lattices and congruences
in the affine gauge}
Let us first express some facts from the theory of quadrilateral 
lattices and discrete congruences in the language of affine geometry which will
be used in the algebraic description of the discrete normal congruences.

Restricting the ambient space from the projective space $\PP^3$ to the
corresponding affine space $\RR^3$, we change from the homogeneous coordinates
$\bx$ to the nonhomogeneous ones $\vbx$, i.e.
$[\bx]=[(\vbx,1)]$. 
Then the Laplace system takes the 
following form \cite{DCN}
\begin{equation} \label{eq:Laplace-dis-aff}
\vbx_{(12)} = A_{(1)} \vbx_{(1)} + 
 B_{(2)}\vbx_{(2)} + (1 -A_{(1)} - B_{(2)})\vbx.
\end{equation}
Define the Lam\'e coefficients $H$ and $G$ as "logarithmic potentials"
\begin{equation}   \label{def:A-H}
A= \frac{H_{(2)}}{H} \; , \quad  B= \frac{G_{(1)}}{G},
\end{equation}
and introduce the suitably scaled tangent 
vectors $\vbX$, $\vbY$,
\begin{equation}  \label{def:HX}
\Delta_1\vbx = H_{(1)} \vbX, \quad \Delta_2\vbx = G_{(2)} \vbY.
\end{equation}
\begin{Cor} \label{cor:freedom-Lame}
The Lam\'e coefficients are not unique; $H$ is given up to the multiplication by a
function of the single argument $m_1$, while $G$ is given up to the multiplication
by a function of the single argument $m_2$
\end{Cor}
In terms of such vectors, the affine Laplace 
equation~\eqref{eq:Laplace-dis-aff} changes into
the system of two equations of the first order
\begin{equation} \label{eq:lin-X}
\Delta_2\vbX = Q_{(2)}\vbY,   \qquad  \Delta_1\vbY = P_{(1)}\vbX , 
\end{equation}
where the proportionality factors $Q$ and $P$, called the rotation coefficients,
 are given by
\begin{equation} \label{eq:lin-H}
 Q = \frac{\Delta_1 G}{ H_{(1)}}, \qquad  
 P = \frac{\Delta_2 H}{ G_{(2)}} .
\end{equation}
 
Denote by $\vby_i$, $i=1,2$, the affine representants of the
focal lattices of a congruence. The coefficients of their 
Laplace equations 
\[
\vby_{i(12)} = A^i_{(1)} \vby_{i(1)} + 
 B^i_{(2)}\vby_{i(2)} + (1 -A^i_{(1)} - B^i_{(2)})\vby_i, \quad i=1,2,
\]
enter into the formulas connecting the lattices as follows~\cite{DCN,TQL} 
\begin{equation} \label{eq:foc-y1-y2}
\vby_2 - \vby_1 = - \frac{1}{B^1-1}\Delta_1\vby_1 = 
\frac{1}{A^2-1}\Delta_2\vby_2.
\end{equation}
Equation \eqref{eq:foc-y1-y2} leads directly to the following result.
\begin{Prop} \label{prop:factorization-y}
Consider a two dimensional congruence whose focal
lattices are represented in the affine gauge by $\vby_i$, $i=1,2$, 
and denote by $H^{i}$, $G^{i}$, their Lam\'e coefficients. 
Define the vectors $\tilde\bX$ and $\tilde\bY$ as follows
\begin{equation}
\tilde\bX = \frac{\vby_2}{H^2}, \qquad \tilde\bY =\frac{\vby_1}{G^1}.
\end{equation}
Then these vectors satisfy the equations 
\begin{equation} 
\Delta_2\tilde\bX = 
 G^1\Delta_2\left(\frac{1}{H^2}\right)  \tilde\bY,  \label{eq:tXY} \qquad 
\Delta_1\tilde\bY = 
 H^2\Delta_1\left(\frac{1}{G^1}\right)  \tilde\bX .  
\end{equation}
\end{Prop}
\begin{Rem}
Notice that the pair $(\frac{1}{H^2},\frac{1}{G^1})$ satisfies the same linear
problem \eqref{eq:tXY} as the pair 
$(\tilde\bX,\tilde\bY)$.
\end{Rem}
\begin{Cor} \label{cor:factorization-y}
Consider a pair $(X^0,Y^0)$ of scalar solutions of the linear problem
\eqref{eq:lin-X}, then
\begin{equation*}
\vby_1 = \frac{\vbY}{Y^0}, \qquad \vby_2 = \frac{\vbX}{X^0},
\end{equation*}
are affine coordinates of the focal lattices of a congruence. Moreover the Lam\'e
coefficients of these lattices can be chosen in such a way that
\[ H^2 = \frac{1}{X^0}, \qquad G^1=\frac{1}{Y^0}.
\]
\end{Cor}

We represent planes in $\RR^3$ by dual vectors $\vbx^*$ such that 
$[\bx^*]=[(\vbx^*,-1)$]. Then the equation of the plane represented by $\vbx^*$ 
and
passing through  the point represented by $\vbx$ is normalized to
\[ \langle \vbx^* , \vbx\rangle = 1 .
\]
\begin{Rem}
Notice that the dual analogue of the points of the plane at infinity
is the set of all planes passing through the origin.
\end{Rem}

In the transition from the homogeneous tangential coordinates of quadrilateral
lattices to the nonhomogeneous ones, the Laplace
equation~\eqref{eq:Laplace-dis-gen-2-dual} is replaced by its affine 
form~\eqref{eq:Laplace-dis-aff}, but with the coefficients 
$A^*$ and $B^*$.
Therefore, also other algebraic considerations, from equations 
\eqref{def:HX} to \eqref{eq:lin-H}, apply to the affine tangential coordinates
of quadrilateral lattices. In this way one defines the dual  Lam\'e
coefficients $H^*$ and $G^*$, the dual normalized tangent vectors $\vbX^*$,
$\vbY^*$, and the dual rotation coefficients $P^*$ and $Q^*$.

The dual version of the connection formulas \eqref{eq:foc-y1-y2} can be
found from the requirement that the four planes represented by 
$\vby^*_1$, $\vby^*_{1(-2)}$, $\vby^*_2$ and $\vby^*_{2(-1)}$ intersect along
one line, exactly in the same way like equations
\eqref{eq:foc-y1-y2} were found from requirement that the four points 
represented by $\vby_1$, $\vby_{1(1)}$, $\vby_2$ and $\vby_{2(2)}$ belong to
one line. Such a derivation, which gives the same result as applying the 
duality principle to equations \eqref{eq:foc-y1-y2} (and taking into account
Corollary~\ref{cor:dual-changes}), leads to the following connection formulas
\begin{equation} \label{eq:foc-y1*-y2*}
\vby^*_{1(-2)} - \vby^*_{2(-1)} = - 
\frac{1}{B^{*2}_{(-1)}-1}\Delta_1\vby^*_{2(-1)} = 
\frac{1}{A^{*1}_{(-2)}-1}\Delta_2\vby^*_{1(-2)},
\end{equation}
which imply the following dual versions of 
Proposition~\ref{prop:factorization-y} and Corollary \ref{cor:factorization-y}. 
\begin{Prop} \label{prop:factorization-y*}
Consider a two dimensional congruence in $\PP^3$ and its focal
lattices whose tangent planes are represented in the affine gauge by 
$\vby_i^*$, $i=1,2$. 
Denote by $H^{*i}$, $G^{*i}$, their Lam\'e coefficients and 
define the dual vectors $\tilde\bX^*$ and $\tilde\bY^*$ as follows
\begin{equation} \label{eq:y12*-XY}
\tilde\bX^* = \left( \frac{\vby^*_1}{H^{*1}}\right)_{(-2)}, \quad 
\tilde\bY^* = \left( \frac{\vby^*_2}{G^{*2}}\right)_{(1)}.
\end{equation}
Then these vectors satisfy the equations 
\begin{equation} 
\Delta_2\tilde\bX^* =  G^{*2}_{(-1)}
 \Delta_2\left(\frac{1}{H^{*1}_{(-2)}}\right)  \tilde\bY^*, \qquad
\Delta_1\tilde\bY^*  = H^{^*1}_{(-2)}
 \Delta_1\left(\frac{1}{G^{*2}_{(-1)}}\right)  \tilde\bX^* .
\end{equation}
\end{Prop}
\begin{Rem}
Notice that the pair $(\frac{1}{H^{*1}_{(-2)}},\frac{1}{G^{*2}_{(-1)}})$ 
satisfies the same linear
problem \eqref{eq:tXY} as the pair 
$(\tilde\bX,\tilde\bY)$.
\end{Rem}
\begin{Cor} \label{cor:factorization-y*}
Given a pair $(X^0,Y^0)$ of scalar solutions of the linear problem
\eqref{eq:lin-X}, then
\[
\vby_1^* =  \left(\frac{\vbX}{X^0}\right)_{(2)}, \qquad
\vby_2^* =  \left(\frac{\vbY}{Y^0}\right)_{(1)},
\]
are affine tangential coordinates of the focal lattices of a congruence.
Moreover the Lam\'e coefficients of the lattices $\vby_1^*$ and $\vby_2^*$ can
be chosen in such a way that
\[ G^{*2} = \frac{1}{Y^0_{(1)}}, \qquad H^{*1}=\frac{1}{X^0_{(2)}}.
\]
\end{Cor}
\subsection{Circular lattices and normal congruences}
In the Euclidean space $\EE^3$ we identify covectors with vectors via the
scalar product and we represent planes using their normals. 
The affine representation of a plane is given by 
$\vbx^*=\vec{\boldsymbol{\eta}} / w$, where
$\vec{\boldsymbol{\eta}}$ is the unit outer (with respect to the origin) 
normal vector to the plane and $w$ is the distance of the plane 
form the origin.

It can be shown \cite{DMS} that a quadrilateral lattice is circular 
(see Section~\ref{sec:normal-cong})
if and only if its normalized tangent
vectors satisfy the constraint 
\begin{equation} \label{eq:circularity1}
\vbX\cdot \vbY_{(1)}+\vbY\cdot \vbX_{(2)} =0  ,
\end{equation}
which, in the continuous limit, gives the orthogonality of the curvature lines.
Other two convenient characterizations of the circular lattice are
as follows (see \cite{KoSchief2,q-red} for more details):
the quadrilateral lattice $x$ is circular if and only if the scalars
\begin{equation} \label{eq:x-X0}
X^\circ:= \frac{1}{2}(\vbx_{(1)} + \vbx)\cdot \vbX \; , \quad 
Y^\circ:= \frac{1}{2}(\vbx_{(2)} + \vbx)\cdot \vbY \; , 
\end{equation}
solve the linear system \eqref{eq:lin-X} or, equivalently, if the
function $|\vbx|^2=\vbx\cdot\vbx$ satisfies
the affine Laplace equation \eqref{eq:Laplace-dis-aff} of $\vbx$.

The algebraic version of the new geometric characterization of circular lattices
described in Proposition~\ref{prop:circ-bis} (we consider here also the
second pair
of bisectrices) is contained in the following
\begin{Prop} \label{prop:circ-alg-new}
For $\vbX$ and $\vbY$ satisfying the linear problem~\eqref{eq:lin-X},
the condition~\eqref{eq:circularity1} is equivalent to the constraint
\begin{equation} \label{eq:circ-alg-new}
\left( \frac{\vbX}{|\vbX|} + \frac{\vbX_{(2)}}{|\vbX_{(2)}|}\right)
\cdot\left(\frac{\vbY}{|\vbY|} + \frac{\vbY_{(1)}}{|\vbY_{(1)}|}\right)=
\left( \frac{\vbX}{|\vbX|} - \frac{\vbX_{(2)}}{|\vbX_{(2)}|}\right)
\cdot\left(\frac{\vbY}{|\vbY|} - \frac{\vbY_{(1)}}{|\vbY_{(1)}|}\right) =0.
\end{equation}
\end{Prop}
\begin{proof}
Equations \eqref{eq:circ-alg-new} are equivalent to equations
\begin{equation}\label{eq:circ-alg-new2}
\frac{\vbX}{|\vbX|}\cdot\frac{\vbY}{|\vbY|}+
\frac{\vbX_{(2)}}{|\vbX_{(2)}|}\cdot\frac{\vbY_{(1)}}{|\vbY_{(1)}|}=
\frac{\vbX}{|\vbX|}\cdot\frac{\vbY_{(1)}}{|\vbY_{(1)}|}+
\frac{\vbX_{(2)}}{|\vbX_{(2)}|}\cdot\frac{\vbY}{|\vbY|}=0.
\end{equation}
($\Rightarrow$) Equation \eqref{eq:circularity1} and the linear problem
\eqref{eq:lin-X} imply~\cite{DMS}, that
\begin{equation} \label{eq:|X|Y|}
\frac{|\vbX_{(2)}|}{|\vbX|}  = \frac{|\vbY_{(1)}|}{|\vbY|}=
\sqrt{1 - Q_{(2)}P_{(1)}}, 
\end{equation}
which together with the following consequence of~\eqref{eq:lin-X}
\begin{equation}
\vbX_{(2)}\cdot\vbY_{(1)} =
(1+P_{(1)}Q_{(2)}) \vbX\cdot\vbY + P_{(1)}|\vbX|^2 + Q_{(2)}|\vbY|^2,
\end{equation}
lead to equations~\eqref{eq:circ-alg-new2}.\\
($\Leftarrow$) Equations~\eqref{eq:circ-alg-new2} and the 
linear problem~\eqref{eq:lin-X} imply~\eqref{eq:|X|Y|} which gives
condition~\eqref{eq:circularity1}.
\end{proof}
\begin{Rem}
Equations \eqref{eq:circ-alg-new2} express the basic property of circular 
quadrilaterals, mentioned in a
proof of Proposition~\ref{prop:circ-bis}, in the form independent of the
convexity
or not of the quadrilateral (see~\cite{DMS}).
\end{Rem}

The following Proposition gives the tangential coordinates of focal lattices of
the normal
congruence of the circular lattice $x$ in terms of point coordinates of the
lattice.
\begin{Prop} \label{prop:dual-normal-congruence}
Given a two dimensional circular lattice $x: \ZZ^2\to\EE^3$,
then the tangent planes of the focal 
lattices of the normal congruence $\nu$ of $x$
are represented in the affine gauge by their normals
\begin{equation} \label{eq:focal-norm-cong}
\vby_1^* = \left(\frac{\Delta_1\vbx}{\Delta_1|\vbx|^2}\right)_{(2)}
= \left(\frac{\vbX}{X^\circ}\right)_{(2)}, \qquad
\vby_2^* =  \left(\frac{\Delta_2\vbx}{\Delta_2|\vbx|^2}\right)_{(1)}
 = \left(\frac{\vbY}{Y^\circ}\right)_{(1)}.
\end{equation}
Moreover the dual Lam\'e coefficients of the lattices $\vby_1^*$ and $\vby_2^*$ can
be chosen in such a way that
\begin{equation} \label{eq:XY0-norm-cong}
G^{*2} = \frac{1}{Y^\circ_{(1)}}, \qquad H^{*1}=\frac{1}{X^\circ_{(2)}}.
\end{equation}
\end{Prop} 
\begin{proof}
The plane $y^*_1$ contains the lines $\nu$ and $\nu_{(2)}$, therefore its
normal must be in the direction of the line passing through the points
$x_{(2)}$ and $x_{(12)}$; i.e. it must be proportional to 
$\Delta_1\vbx_{(2)}$.
The normalization factor can be found from 
the condition that the plane passes through the middle-point 
of the segment $x_{(2)} x_{(12)}$.
Similarly we find the expression for $\vby_2^*$. 
The rest of the Proposition follows from Corollary~\ref{cor:factorization-y*}.
\end{proof}
Let us give the algebraic characterization, modeled
on Proposition~\ref{prop:dual-normal-congruence}, of the normal congruences.
\begin{Prop} \label{prop:char-norm-cong}
Consider a two dimensional congruence in the Euclidean space, with the focal
lattices represented in the affine gauge by their normals $\vby_i^*$, $i=1,2$, 
and denote by $H^{*i}$, $G^{*i}$, their dual Lam\'e coefficients. 
The congruence is normal if and only if
\begin{equation} \label{eq:dual-normal-congruence-cond}
\left(\frac{\vby_{1}^*}{H^{*1}}\right)_{(-2)}\cdot 
\left(\frac{\vby_{2}^*}{G^{*2}}\right) +
\left(\frac{\vby_{2}^*}{G^{*2}}\right)_{(-1)}\cdot 
\left(\frac{\vby_{1}^*}{H^{*1}}\right) =0.
\end{equation}
\end{Prop}
\begin{proof}
Define the vectors $\vbX$ and $\vbY$ by equation~\eqref{eq:y12*-XY};
then the condition \eqref{eq:dual-normal-congruence-cond} is transformed into 
equation~\eqref{eq:circularity1}. 
By Proposition~\ref{prop:factorization-y*} the vectors
$\vbX$, $\vbY$ satisfy the linear problem of the form~\eqref{eq:lin-X}, 
with the rotation coefficients $P$, $Q$ given by
\begin{equation}
P_{(-1)} = H^{^*1}_{(-2)} \Delta_1\left(\frac{1}{G^{*2}_{(-1)}}\right), \qquad
Q_{(-2)} = G^{*2}_{(-1)} \Delta_2\left(\frac{1}{H^{*1}_{(-2)}}\right).
\end{equation}
The vectors
\begin{equation*}
\vB_{1\pm} = \frac{\vbX}{|\vbX|} \pm \frac{\vbX_{(2)}}{|\vbX_{(2)}|}
\end{equation*}
are normal vectors of two orthogonal
planes bisecting $y^*_{1(-2)}$ and $y^*_{1}$, 
similarly the vectors
\begin{equation*}
\vB_{2\pm} = \frac{\vbY}{|\vbY|} \pm \frac{\vbY_{(1)}}{|\vbY_{(1)}|}
\end{equation*}
are normal vector of the planes bisecting $y^*_{2(-1)}$ and $y^*_{2}$.
The rest of the proof follows from Proposition~\ref{prop:circ-alg-new}.
\end{proof} 
\begin{Prop}
Given a discrete normal congruence $\nu$, then there exists a circular lattice 
$x$ such that $\nu$ is the normal congruence of $x$.
\end{Prop}
\begin{proof}
Define the vectors $\vbX$, $\vbY$ and their rotation coefficients $P$, $Q$ like
in the proof of Proposition~\ref{eq:dual-normal-congruence-cond}. 
By the remark after Proposition~\ref{prop:factorization-y*} 
the functions $X^\circ$, $Y^\circ$, defined by
equation~\eqref{eq:XY0-norm-cong},
satisfy the same linear problem as $\vbX$, $\vbY$.

Consider the following linear system for $\vbx$
\begin{align} \label{eq:construction-x1}
\vbx_{(1)}  = & \vbx + \frac{2(X^\circ-\vbx\cdot\vbX)}{|\vbX|^2}\vbX, \\
\label{eq:construction-x2}
\vbx_{(2)}  = & \vbx + \frac{2(Y^\circ-\vbx\cdot\vbY)}{|\vbY|^2}\vbY,
\end{align}
whose geometric interpretation was described at the end of
Section~\ref{sec:normal-cong}. 
Its compatibility is asserted by 
equations \eqref{eq:lin-X}, \eqref{eq:circularity1} and their 
consequence \eqref{eq:|X|Y|}.

The functions $H$ and $G$ defined by
\begin{eqnarray*}
H_{(1)} & = & \frac{2(X^\circ-\vbx\cdot\vbX)}{|\vbX|^2}, \\
G_{(2)} & = & \frac{2(Y^\circ-\vbx\cdot\vbY)}{|\vbY|^2},
\end{eqnarray*}
satisfy equations \eqref{eq:lin-H} and are connected with $\vbx$ and the pair
$(\vbX,\vbY)$ by equations~\eqref{def:HX}. Therefore the lattice represented by
$\vbx$ is a quadrilateral lattice with the normalized tangent vectors $\vbX$
and $\vbY$, and the Lam\'e coefficients $H$ and $G$. Because $\vbX$
and $\vbY$ satisfy the constraint \eqref{eq:circularity1}, the
lattice is circular.

Equations~\eqref{eq:construction-x1}-\eqref{eq:construction-x2} imply
that the circular lattice $x$ and the focal lattices $y_i$
of the congruence are connected by equations~\eqref{eq:focal-norm-cong};
which means just that the focal planes bisect orthogonally corresponding segments of
the lattice, i.e. the congruence $\nu$ is the normal congruence of $x$.   
\end{proof}
\begin{Rem}
If there exists one such circular lattice, then there exist an infinity of such
circular lattices labelled by the position of an initial point of the
construction.
\end{Rem}

\section{The Darboux transformation for the discrete Bianchi system}
The following results can be checked by direct (but tedious) calculation
and actually are modification of Theorem~7 of \cite{DNS-I}.
\begin{Lem} \label{lem:lin-sys-B}
Given $\vbn:\ZZ^2\to\EE^3$ satisfying the discrete Bianchi 
system~\eqref{eq:Bianchi-Koenigs-disc}, define $\vecbfeta_0=F\vbn$ and let 
$\vecbfeta_1$ and $\vecbfeta_2$ be unit vectors
orthogonal to $\vecbfeta_0$ and to each other. If $p^B_A$, $q^B_A$,
$A,B=0,1,2$, are the functions defined by the unique decompositions
\[ \vecbfeta_{A} = \sum_{B=0}^2 p^B_A \vecbfeta_{B(1)}, \quad
\vecbfeta_{A} = \sum_{B=0}^2 q^B_A \vecbfeta_{B(2)},
\]
and if 
\[R(m_1,m_2) = U_1(m_1)+U_2(m_2), \quad a(m_1)=U_1(m_1) +\lambda, \quad 
b(m_2)=U_2(m_2) -\lambda,
\]
then:\\
(i) the linear system
\begin{equation} \label{eq:lin-BE}
\Theta_{(i)}=M_i\Theta, \qquad i=1,2,
\end{equation}
where $\Theta=(\theta, \theta^\prime, \theta^{\prime\prime}, y^1, y^2)^T$, and 
\begin{align*}
M_1 &= \left( \begin{array}{ccccc}
0&1&0&0&0 \\
\frac{R_{(1)}\frac{p_0^0}{F}-b}{a_{(1)}}&
\frac{b F- R_{(1)}(\frac{R+b}{R}p_0^0-\frac{1}{F _{(1)}})}{a_{(1)}}&
\frac{b}{a_{(1)}}(F-\frac{R_{(1)}}{R}p_0^0)&
\; \; \frac{R_{(1)}}{a_{(1)}}p_{1}^0& \; \;\frac{R_{(1)}}{a_{(1)}}p_{2}^0 \\
-1&F&F&0&0\\
\frac{p_0^{1}}{F} & -\frac{R+b}{R}p_0^{1} &
-\frac{b}{R}p_0^{1} & p_{1}^{1} & p_{2}^{1}\\
\frac{p_0^{2}}{F} & -\frac{R+b}{R}p_0^{2} &
-\frac{b}{R}p_0^{2} & p_{1}^{2} & p_{2}^{2}
\end{array}\right), \\
M_2 &= \left( \begin{array}{ccccc}
0&0&1&0&0\\
-1&F&F&0&0\\
\frac{R_{(2)}\frac{q_0^0}{F}-a}{b_{(2)}}&
\frac{a}{b_{(2)}}(F-\frac{R_{(2)}}{R}q_0^0)&
\frac{a F- R_{(2)}(\frac{R+a}{R}q_0^0-\frac{1}{F _{(2)}})}{b_{(2)}}&
-\frac{R_{(2)}}{b_{(2)}}q_{1}^0&-\frac{R_{(2)}}{b_{(2)}}q_{2}^0
\\
-\frac{q_0^{1}}{F}&\frac{a}{R}q_0^{1}&\frac{a+R}{R}q_0^{1}
&q_{1}^{1}&q_{2}^{1}\\
-\frac{q_0^{2}}{F}&\frac{a}{R}q_0^{2}&\frac{a+R}{R}q_0^{2}
&q_{1}^{2}&q_{2}^{2}
\end{array}\right),
\end{align*}
is compatible;\\
(ii) the function
\begin{equation*} \label{eq:constraint}
I = (y^1)^2+(y^2)^2 +\frac{R}{F}\theta^2 +
\frac{F}{R}\left(
b\theta^{\prime\prime}-a\theta^\prime\right)^2 -
2\theta\left( a\theta^\prime +b\theta^{\prime\prime} \right),
\end{equation*}
is a first integral of the system;\\
(iii) the function $\vbomega=\sum_{A=0}^2 y^A\vecbfeta_A$ with 
\begin{equation}
y^0 = \frac{ a\theta^\prime - b\theta^{\prime\prime}}{R},
\end{equation}
satisfies the linear system
\eqref{eq:lin-w-1}-\eqref{eq:lin-w-2} with $\vbn$ in place of $\bx^*$. 
\end{Lem}
\begin{Cor}
Notice that the solution $\theta$ of the linear system 
\eqref{eq:lin-BE}
is a solution of the discrete Moutard equation \eqref{eq:Moutard-dis}, while
$\theta^\prime=\theta_{(1)}$ and $\theta^{\prime\prime}=\theta_{(2)}$.
\end{Cor}
\begin{Th}
Given $\vbn:\ZZ^2\to\EE^3$ satisfying the discrete Bianchi 
system~\eqref{eq:Bianchi-Koenigs-disc}, let $\theta$ and $\vbomega$
be obtained from the
solution of the linear system \eqref{eq:lin-BE} 
subjected to the admissible constraint $I=0$. Then 
$\vbomega$ satisfies equations \eqref{eq:w*n}-\eqref{eq:w*w} and $\vbn^\prime$,
given by equation~\eqref{eq:tang-x'} with $\vbn$ and $\vbomega$ in place of 
$\bx^*$ and $\bw^*$, is a new solution of the discrete Bianchi 
system~\eqref{eq:Bianchi-Koenigs-disc}.
\end{Th}
\bibliographystyle{amsplain}

\providecommand{\bysame}{\leavevmode\hbox to3em{\hrulefill}\thinspace}
\providecommand{\MR}{\relax\ifhmode\unskip\space\fi MR }
\providecommand{\MRhref}[2]{%
  \href{http://www.ams.org/mathscinet-getitem?mr=#1}{#2}
}
\providecommand{\href}[2]{#2}

\end{document}